\documentclass[twocolumn,showpacs,superscriptaddress,amsmath,amssymb]{revtex4}
\usepackage{graphicx}
\usepackage{dcolumn}
\usepackage{bm}

\def\bd{
\begin{document}} \def\ed{\end{document}}
\def\bmp{\begin{minipage}} \def\emp{\end{minipage}}
\def\bcc{\begin{center}} \def\ecc{\end{center}}     \def\npg{\newpage}
\def\beq{\begin{equation}} \def\eeq{\end{equation}} \def\hph{\hphantom}
\def\be{\begin{equation}} \def\ee{\end{equation}} \def\r#1{$^{[#1]}$}
\def\n{\noindent} \def\ni{\noindent} \def\pa{\parindent}
\def\hs{\hskip} \def\vs{\vskip} \def\hf{\hfill} \def\ej{\vfill\eject}
\def\cl{\centerline} \def\ob{\obeylines}  \def\ls{\leftskip}
\def\underbar#1{$\setbox0=\hbox{#1} \dp0=1.5pt \mathsurround=0pt
   \underline{\box0}$}   \def\ub{\underbar}    \def\ul{\underline}
\def\f{\left} \def\g{\right} \def\e{{\rm e}} \def\o{\over} \def\d{{\rm d}}
\def\vf{\varphi} \def\pl{\partial} \def\cov{{\rm cov}} \def\ch{{\rm ch}}
\def\la{\langle} \def\ra{\rangle} \def\EE{e$^+$e$^-$} \def\pt{p_{\rm t}}
\def\pti{p_{{\rm t},i}} \def\ptj{p_{{\rm t},j}}
\def\bitz{\begin{itemize}} \def\eitz{\end{itemize}}
\def\btbl{\begin{tabular}} \def\etbl{\end{tabular}}
\def\btbb{\begin{tabbing}} \def\etbb{\end{tabbing}}
\def\beqar{\begin{eqnarray}} \def\eeqar{\end{eqnarray}}
\def\\{\hfill\break} \def\dit{\item{-}} \def\i{\item}
\def\bbb{\begin{thebibliography}{9} \itemsep=-1mm}
\def\ebb{\end{thebibliography}} \def\bb{\bibitem}
\def\bpic{\begin{picture}(260,240)} \def\epic{\end{picture}}
\def\akgt{\cl{\bf ACKNOWLEDGMENTS}}
\def\fgn{\noindent{\bf\large\bf figure captions}}
\def\lan{\langle}
\def\ran{\rangle}
\def\p{\pi}
\def\ifmath#1{\relax\ifmmode #1\else $#1$\fi}%
\def\rc{\ifmath{{\mathrm{c}}}}
\def\cut{\ifmath{{\mathrm{cut}}}}
\def\rF{\ifmath{{\mathrm{F}}}}
\def\rK{\ifmath{{\mathrm{K}}}}
\def\rp{\ifmath{{\mathrm{p}}}}
\def\rt{\ifmath{{\mathrm{t}}}}
\def\LAB{\ifmath{{\mathrm{LAB}}}}
\def\cut{\ifmath{{\mathrm{cut}}}}
\def\beq{\begin{equation}}
\def\eeq{\end{equation}}

\newcommand{\cinst}[2]{$^{\mathrm{#1}}$~#2\par}
\newcommand{\crefi}[1]{$^{\mathrm{#1}}$}
\newcommand{\crefii}[2]{$^{\mathrm{#1,#2}}$}
\newcommand{\crefiii}[3]{$^{\mathrm{#1,#2,#3}}$}
\newcommand{\HRule}{\rule{0.5\linewidth}{0.5mm}}

\bd
\title{ Locating critical point of QCD phase transition\\ basing on finite-size scaling}

\author{Chen Lizhu}
\affiliation{Institute of Particle Physics, Hua-Zhong Normal
University, Wuhan 430079, China}
\author{X.S. Chen} \affiliation{Institute of Theoretial
Physics, Chinese Academy of Sciences, Beijing 100190, China}
\author{Wu Yuanfang} \affiliation{Institute of Particle Physics, Hua-Zhong
Normal University, Wuhan 430079, China}\affiliation{Key Laboratory
of Quark $\&$ Lepton Physics (Huazhong Normal University),
Ministry of Education, China }

\begin{abstract}
It is argued that in relativistic heavy ion collisions, due to
limited size of the formed matter, the reliable criterion of
critical point is finite-size scaling, rather than non-monotonous
behavior of observable. How to locate critical point by
finite-size scaling is proposed. The data of $\pt$ correlation
from RHIC/STAR are analyzed. Critical points are likely observed
around $\sqrt s =62$ and $200$ GeV. They could be, respectively,
the transition of deconfinement and chiral symmetry restoration
predicted by lattice-QCD. Further confirmation with other
observable and energies is suggested.
\end{abstract}


\maketitle Lattice-QCD simulations have shown that the transition
of deconfinement in quantum chromodynamics (QCD) at vanishing
baryon chemical potential $\mu_{\rm B}$ is
crossover~\cite{lattice-1}. There has been much speculation that
the crossover becomes a true first-order phase transition for
larger values of $\mu_{\rm B}$. This suggests that the QCD phase
diagram can exhibit a critical endpoint where the line of first
order transition matches that of second order or analytical
crossover~\cite{1st}.

Chiral symmetry restoration is another QCD originated phase
transition. It has been shown that the transition for $\mu_{\rm
B}=0$ is crossover~\cite{c-crossover}. So there could also be a
chiral critical endpoint in phase diagram. But it is unclear if
the critical temperature of chiral symmetry restoration is
above~\cite{karsch-h}, or equal to~\cite{karsch-s}, or
below~\cite{quarkyonic} that of the deconfinement.

Locating the critical endpoints of QCD phase transitions by
lattice calculation is still a formidable challenge. But if the
critical endpoint is in the region accessible to current
relativistic heavy ion collisions, it should be discovered
experimentally.

Most of the current signatures for finding the critical point are
focused on the anomalous, or non-monotonous, behavior of the
observable at various incident energies~\cite{ebye-t}. The
argument is that in infinite system, the correlation length $\xi$
diverges when approaching the critical point. The contribution of
this singularity to the observable is supposed to be proportional
to $\xi^2$. However, the data from RHIC and SPS in more than a
decade accumulation show no sign of anomalous behavior as a
function of $\sqrt s$~\cite{ebye-e}.

In relativistic heavy ion collisions, two nuclei move with
relativistic velocity and collide as two contracted pancakes. More
central collision makes overlapped area larger. It is just because
the large number of strongly interacting nucleons in more central
nuclear collisions make the transition between hadron and
quark-gluon plasma possible. The centrality (or the system size)
dependence of the observable is noticeable~\cite{starprc}.

Due to the finite size of system, no divergence can be practically
observed at critical point. The physical quantities, which are
divergent in infinite system, become finite and have a maximum,
i.e., so called non-monotonous behavior. However, the position of
the maximum changes with system size and deviates from the true
critical point.

The appearance of non-monotonous behavior is not always associated
with critical point. Taking one-dimensional Ising model as an
example, there is no critical point in this model, but its
specific heat in a finite system has non-monotonous behavior.

Moreover, the absence of non-monotonous behavior does not mean no
critical point. The physical quantities like order parameter,
which are finite in infinite system, have a monotonous behavior
near critical point in a finite system~\cite{chen1996}. Therefore,
non-monotonous behavior is not a reliable criterion for the
critical point of finite system.

An effective identification of critical point of finite system is
the finite-size scaling, which was proposed from
phenomenological~\cite{fss-1} and
renormalization-group~\cite{fss-RG} theories, and was approved by
the Monte Carlo results of finite systems in different universal
classes~\cite{fss-2}.

In this letter, we first propose how to locate critical point by
finite-size scaling. Then the data of $\pt$ correlation at 6
centralities and 4 incident energies from RHIC/STAR are analyzed.
The behavior of critical point is likely observed around $\sqrt s
= 62$ and $200$ GeV. Finally, we suggest how to confirm the
findings and precisely locate the critical point in coming
experimental study.

The main points of finite-size scaling can be described as the
following. An observable $Q$ of finite system is a function of
temperature $T$ and system size $L$. When $L$ is much larger than
the microscopic length scale and $T$ is in the vicinity of
critical point $T_c$, the observable $Q(T,L)$ can be written in a
finite-size scaling form~\cite{fss-1,fss-RG,fss-2},
\begin{equation}\label{efss}
 Q(T,L)=L^{\lambda/\nu}F_Q(tL^{1/\nu}).
 \end{equation}
\noindent $t=(T-T_c)/T_c$ is the reduced temperature and $\lambda$
is the critical exponent of the observable. $\nu$ is the critical
exponent of the correlation length $\xi =\xi_0 t^{-\nu}$.

Finite-size scaling not only characterizes the scaling behavior of
thermodynamic quantities of finite system near critical point, but
also provides criterion for locating the critical point. At
critical point $T=T_c$, the finite-size scaling function $F_Q$ in
Eq.~(\ref{efss}) becomes
\begin{equation}\label{efss-1}
F_Q (0) = Q(T_c,L)L^{-\lambda/\nu},
\end{equation}
\noindent which is constant and independent of system size $L$. In
the plot of $ Q(T,L)L^{-\lambda/\nu}$ {\it vs} $T$, the critical
point [$T_c$, $F_Q (0)$] is a {\it fixed point}, where all curves
of different system sizes converges to. Reversely, the appearance
of fixed point indicates the existence of a critical point.

If the critical exponent $\lambda =0$, like Binder cumulant
ratio~\cite{binder1981}, the fixed point can be obtained directly
from the temperature dependence of this observable at different
system sizes. This is why Binder cumulant ratio has been used very
widely in determining critical point of finite-size system.

If the critical exponent $\lambda\neq 0$ and is unknown, the fixed
point can be found by investigating the temperature dependence of
$Q(T,L)L^{-a}$ at different system sizes. When a fixed point is
observed at a certain parameter $a_0$, it indicates the existence
of a critical point and the parameter $a_0$ is related to the
ratio of critical exponents, i.e., $\lambda/\nu=a_0$.

The critical point can also be found directly from the system size
dependence of the observable. Taking logarithm in the both sides
of Eq.~(\ref{efss}), it becomes
\begin{equation}\label{efss-2}
\ln Q(T,L)= \lambda/\nu \ln L + \ln F_Q (tL^{1/\nu}).
\end{equation}
At critical point $t=0$, the second term of Eq.~(\ref{efss-2})
becomes a constant and $\ln Q(T_c,L)$ becomes a straight line with
respect to $\ln L$. If system is away from the critical point, the
second term of Eq.~(\ref{efss-2}) is no longer a constant. It
gives an additional size dependent contribution to the observable
and makes $\ln Q(T,L)$ deviate from the straight line with respect
to $\ln L$.

It is found recently that the finite-size scaling holds not only
for thermodynamic quantities like order-parameter, susceptibility,
and so on, but also for various cluster sizes~\cite{liangsheng}
and their fluctuations~\cite{lizhu-ising}. Therefore, the
finite-size scaling of various critical related observable could
be used to identify critical point and its critical exponents.

In relativistic heavy ion collision, correlation and fluctuation
of final state particles is regarded as critical related
observable~\cite{lattice-corr}. Although much attention have been
drawn in measuring them, but influence of system size has been
neglected. The available data for system size study is very few.
The $\pt$ correlation at Au+ Au collisions from
RHIC/STAR~\cite{starprc} is the only data which can be used for
the analysis, where the centrality dependence of $\pt$ correlation
at 4 incident energies are well presented~\cite{starprc}. But the
errors of the data at $\sqrt s=20$ GeV are much larger than that
at other collision energies. The $\pt$ correlation is defined as
\beqar\label{ept} P(\sqrt s, L)=\frac{1}{N_{\rm
e}}\sum\limits_{k=1}^{N_{\rm
e}}\frac{\sum\limits_{i=1}^{N_k}\sum\limits_{j=1,i\not=j}^{N_k}(\pti-\la
\pt\ra)(\ptj-\la\pt\ra)}{N_k(N_k-1)}. \eeqar

\noindent $N_{\rm e}$ is the number of event, $\pti$ is the
transverse-momentum of the $i$th particle in each event, and $N_k$
is the number of particles in the $k$th event. $\la\ldots\ra$ is
the average over event sample.

Collision energy is the controllable condition. Here we let it
play the role of temperature in the analysis of finite-size
scaling. The size of the formed matter is mainly limited by the
size of overlapping transverse region, which is proportional to
the number of participant nucleons and is quantified as
centrality. So the initial mean size of the formed matter can be
approximately estimated by the square root of participants, $\sqrt
{N_{\rm part}}$. We choose dimensionless (or relative) size,
\beqar\label{el} L=\sqrt {N_{\rm part}}/\sqrt {2N_{\rm A}},\eeqar

\noindent as scaled mean size of initial system, where $N_{\rm A}$
is the number of nucleons of incident nucleus. The system size at
transition should be a monotonically increasing function of $L$.
The position of critical point is insensitive to the concrete form
of this function, but only the critical exponents changes with it.
As the first step, the system size at transition is assumed to be
proportional to $L$.

In the case of a few critical points, the finite-size scaling of
$\pt$ correlation in the vicinity of each critical collision
energy $\sqrt{s_{c,i}} (i=1,2,...)$ can be written as
\beqar\label{ptfss} P(\sqrt
s,L)=L^{\lambda_i/\nu_i}F_{P,i}[e_iL^{1/\nu_i}]. \eeqar

\noindent $e_i=(\sqrt{s}-\sqrt {s_{c,i}})/\sqrt {s_{c,i}}$ is the
reduced collision energy at $i$th critical point, which is unknown
in priori. $\lambda_i$ is the $i$th critical exponent of $\pt$
correlation. In the following, we demonstrate how to locate the
critical point by the data of $\pt$ correlation from RHIC/STAR.

\begin{figure*}
\includegraphics[width=7.in]{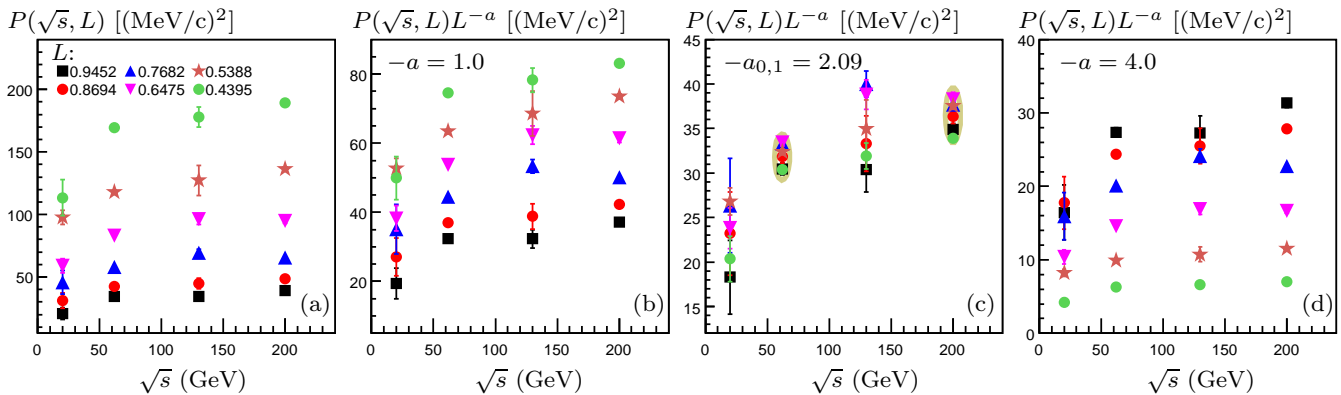}
\caption{\label{Fig. 1} (a) The energy dependence of $\pt$
correlations at different sizes $L$ (or centralities). Data come
from RHIC/STAR~\cite{starprc}. (b), (c) and (d) are $\pt$
correlation multiplied by the factor, $L^{-a}$, with $-a=1.0$,
2.09 and 4, respectively}.
\end{figure*}

Firstly, we change the centrality dependence of $\pt$ correlation
at different collision energies in Ref.~\cite{starprc} to the
collision energy dependence at different sizes (or centralities).
The results are shown in Fig.~1(a). Since in the most peripheral
collisions, the size of the formed matter is too small to be
inside the asymptotic region of finite-size scaling, we choose six
centralities at mid-central and central collisions to do the
analysis. The sizes corresponding to the 6 centralities are
indicated in the legend of Fig.~1(a). It is clear that at a given
collision energy, the correlation strength increases with the
decrease of system size. The influence of finite size is obvious.

If critical collision energy of QCD phase transition is in the
range of incident energy at RHIC, the behavior of fixed point
should be observable. So we multiply $P(\sqrt s, L)$ by a size
factor $L^{-a}$ with different $a$ to see how it changes with the
system size $L$. Varying $-a$ from small to large, it is
interesting to see that at collision energy $\sqrt s =62$ GeV, all
points of different sizes move firstly toward each other, then
well converge at $-a_{0,1}=2.09$, and finally move again apart
from each other. The corresponding steps and typical $a$ values
are presented in Fig.~1(b), (c), and (d) respectively, where the
errors in each sub-figures come from the measure of $P(\sqrt s,
L)$ only, and the errors of $N_{\rm part}$ are not included.

At $\sqrt s = 200$ GeV, the points of different sizes show the
same behavior and best converge at $-a_{0,2}=2.08$. While in the
whole process, the points of different sizes at energies $\sqrt s
= 20$ (or 130) GeV never move close to each other as those at
$\sqrt s =62$ (or 200) GeV do. So there are likely two fixed
points around $\sqrt s =62$ and 200 GeV.

In order to confirm the position of fixed points, we study the
$\ln L$ dependence of $\ln P(\sqrt s, L)$ for four incident
energies, respectively. A parabola fit, $c_2(\ln L)^2+c_1\ln
L+c_0$, is used at each collision energy. The better straight-line
behavior results in smaller $|c_2|$ and larger ratio of
$|c_1/c_2|$. The fit parameters, $c_2$ and $c_1$, for 4 collision
energies are listed in Tab.~1. It shows that the better
straight-line behavior happen to be at $\sqrt s =62$ and 200 GeV,
which are the same collision energies of fixed points found above.
The data at these two energies can be well fitted, respectively,
by the straight lines with slopes $a_{0,1}$ and $a_{0,2}$ obtained
above by the fixed points. The results are shown in Fig.~2(a).
While, the data at $\sqrt s = 20$ and 130 GeV are better fitted by
parabola as shown in Fig.~2(b).

{\small
\begin{table}
\caption{\label{Table 1.} Parameters of parabola fits.}
\begin{tabular}{ccccccccc}\hline
  $\sqrt s $(GeV) && 20 && 62 && 130 && 200 \\ \hline
  $ |c_2|$ && 1.86$\pm$ 0.93 && 0.6 $\pm$ 0.09 && 1.56$\pm$ 0.41  && 0.77$\pm$ 0.1 \\
 \hline
$ |c_1|$ && 3.9$\pm$0.89 && 2.59$\pm$ 0.09 && 3.43$\pm$ 0.41 &&
2.74$\pm$0.1 \\ \hline
\end{tabular}
\end{table}}


\begin{figure}
\includegraphics[width=3.2in]{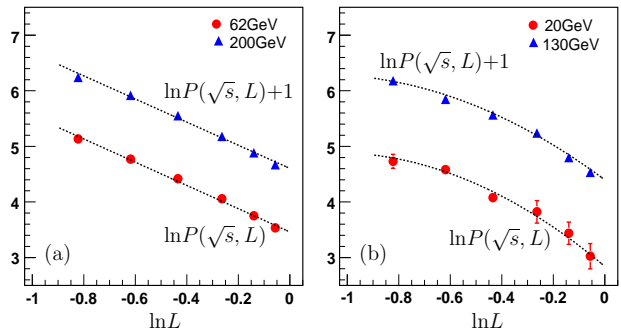}
\caption{\label{Fig. 1} Double-log plots of $\pt$ correlation with
respect to size, (a): straight-line fits with slopes $a_{0,1}$ and
$a_{0,2}$ obtained by fixed points, and (b): parabola fits. }
\end{figure}

The same analysis has also been applied to the $\pt$ correlation
normalized by the average $\pt$ over the whole
sample~\cite{starprc}. The analysis for normalized $\pt$
correlation at $\sqrt s =62$ and $200$ GeV show exactly the same
behavior of fixed points and straight lines as what $\pt$
correlation demonstrates above. The critical exponents of
normalized $\pt$ correlation (about 1.1) are smaller than that of
$\pt$ correlation.

So the critical collision energies are most probably around $\sqrt
s =62$ and $200$ GeV, rather than near $\sqrt s = 20$ and 130 GeV.
The same analysis for other critical related observable, such as
the fluctuations of mean $\pt$ per event, the moments of
multiplicity, the ratio of $K$ to $\pi$, and so on, will be
greatly helpful in confirming the observed results. Therefore, the
incident energy and centrality dependence of those observable are
called for.

If there were additional collisions around $\sqrt s =62$ and $200$
GeV, we could determine the finite-size scaling function defined
in Eq.~(\ref{ptfss}). This is impossible at present since there
are only two collision energies in addition to the critical ones,
and they could be outside of the asymptotic region where
finite-size scaling holds.

The findings of the two critical points may imply that
deconfinement and chiral symmetry restoration occur at different
temperatures. Which one is at the lower or higher temperature
(energy) has to be confirmed finally from theoretical calculation.
Two critical collision energies, $\sqrt s =62$ and $200$ GeV, are
both within the range estimated by lattice calculation~\cite{hTc}.

The similar ratios of critical exponents at two critical points is
consistent with current theoretical estimation, which shows that
all critical exponents of the deconfinement transition, in the
same university as the $3$-dimensional Ising
model~\cite{3d-ising}, are very close to that of chiral symmetry
restoration, in the same university as the $3$-dimensional O$(4)$
model with spin symmetry~\cite{3d-o4}.

To the summary, we argue in this letter that finite-size effects
of the formed matter in relativistic heavy ion collisions is not
negligible. The finite-size scaling, rather than non-monotonous
behavior of observable is a reliable criterion of the existence of
critical point. Then we propose how to locate critical point by
finite-size scaling. As an application, we analyze the data of
$\pt$ correlation and its normalized one at 6 centralities and 4
incident energies from RHIC/STAR. Two fixed points, and therefore
two critical points, are likely observed around $\sqrt s = 62$ and
200GeV. They could be, respectively, related to the transition of
deconfinement and chiral symmetry restoration predicted by
lattice-QCD. The ratios of critical exponents at these two
critical points are similar, in consistence with current
theoretical estimation.

The confirmation of this observation requires the efforts from
both theoretical and experimental sides. From experimental side,
it is proposed to get more and better data on other critical
related observable at current collision energies, and a few
additional collisions around $\sqrt s = 62$ and 200 GeV. Then we
can more precisely determine the critical endpoints and critical
exponents.

The authors are grateful to Dr. Li Liangsheng, Prof. Liu Lianshou
and Prof. Dr. Hou Defu for very helpful discussions. This work is
supported in part by the NSFC of China with project No. 10835005
and MOE of China with project No. IRT0624 and No. B08033.

\ed